\pdfoutput=1
\documentclass[useAMS,usenatbib]{mnras}
\usepackage{graphicx}
\usepackage{amsbsy,amsmath}
\usepackage{graphics,graphicx}
\usepackage{subfloat}
\usepackage[compatibility=false]{caption}
\usepackage[skip=0.1pt]{subcaption}
\usepackage{ae,aecompl}
\usepackage[compact]{titlesec}
\usepackage{notoccite}
\usepackage{caption}
\usepackage{subcaption}
\usepackage{amsbsy,amsmath}
\usepackage{graphics,graphicx}
\usepackage{longtable}
\usepackage[latin1]{inputenc}
\usepackage{amsmath}
\usepackage{amsfonts}
\usepackage{amssymb}
\usepackage{makeidx}
\usepackage{float}
\restylefloat{figure}
\usepackage{stfloats}
\usepackage{epsfig}
\usepackage{graphicx}
\usepackage{caption}
\usepackage{color}
\usepackage{tabularx}
\usepackage{natbib}
\usepackage{multicol}
\usepackage{gensymb}
\usepackage[toc]{appendix}
\usepackage{breqn}

\title{MeerKAT's discovery of a radio relic in the bimodal merging cluster A2384 }

\author[Parekh et al.]{V. Parekh$^{1,2}$\thanks{E-mail: viralp@ska.ac.za}, 
K. Thorat$^{3}$, 
R. Kale$^{4}$, 
B. Hugo$^{2}$, 
N. Oozeer$^{2,8}$,
S. Makhathini$^{1}$,
D. Kleiner$^{5}$, \newauthor
S. V. White$^{1}$,
G. I. G. J\'ozsa$^{1,2,6}$,
O. Smirnov$^{1,2}$,
K. van der Heyden$^{7}$,
S. Perkins$^{2}$,\newauthor
L. Andati$^{1}$,
A. Ramaila$^{2}$, 
and M. Ramatsoku$^{1}$ \\
$^{1}$Department of Physics and Electronics, Rhodes University, PO Box 94, Makhanda, 6140, South Africa\\
$^{2}$South African Radio Astronomy Observatory, 2 Fir Street, Black River Park, Observatory, Cape Town, 7925, South Africa\\
$^{3}$ University of Pretoria, Lynnwood Rd, Hatfield, Pretoria, 0002, South Africa\\
$^{4}$ National Centre for Radio Astrophysics, Savitribai Phule Pune University Campus, Ganeshkhind, Pune, Maharashtra 411007, India\\ 
$^{5}$ INAF - Osservatorio Astronomico di Cagliari, Via della Scienza 5, I-09047 Selargius (CA), Italy\\
$^{6}$ Argelander-Institut f\" ur Astronomie, Auf dem H\" ugel 71, D-53121 Bonn, Germany\\
$^{7}$ Department of Astronomy, R W James Building, University of Cape Town, Rondebosch, Cape Town, 7700, Republic of South Africa \\
$^{8}$ African Institute for Mathematical Sciences, 6 Melrose Road, Muizenberg, 7945, South Africa \\
}

\date{Accepted XXX. Received YYY; in original form ZZZ}

\pubyear{2015}

\begin{document}
\label{firstpage}
\pagerange{\pageref{firstpage}--\pageref{lastpage}}
\maketitle

\begin{abstract}
We present the discovery of a single radio relic located at the edge of the galaxy cluster A2384, using the MeerKAT radio telescope. A2384 is a nearby ($z$ = 0.092), low mass, complex bimodal, merging galaxy cluster that displays a dense X-ray filament ($\sim$ 700 kpc in length) between A2384(N) (Northern cluster) and A2384(S) (Southern cluster). The origin of the radio relic is puzzling. By using the MeerKAT observation of A2384, we estimate that the physical size of the radio relic is 824 $\times$ 264 kpc$^{2}$ and that it is a steep spectrum source. The radio power of the relic is $P_{1.4\mathrm{GHz}}$ $\sim$ (3.87 $\pm$ 0.40) $\times$ 10$^{23}$ W Hz$^{-1}$. This radio relic could be the result of shock wave propagation during the passage of the low-mass A2384(S) cluster through the massive A2384(N) cluster, creating a trail appearing as a hot X-ray filament. In the previous GMRT 325 MHz observation we detected a peculiar FR I radio galaxy interacting with the hot X-ray filament of A2384, but the extended radio relic was not detected; it was confused with the southern lobe of the FR I galaxy. This newly detected radio relic is elongated and perpendicular to the merger axis, as seen in other relic clusters. In addition to the relic, we notice a candidate radio ridge in the hot X-ray filament. The physical size of the radio ridge source is $\sim$ 182 $\times$ 129 kpc$^{2}$. Detection of the diffuse radio sources in the X-ray filament is a rare phenomenon, and could be a new class of radio source found between the two merging clusters of A2384(N) and A2384(S).
\end{abstract}

\section{Introduction}
\par \par In the large scale structure (LSS) formation process, galaxy clusters are formed at the nodes of complex cosmic web filaments \citep{2005Natur.435..629S,2008SSRv..134..311D,2014MNRAS.438.3465T,2017ApJ...844...25B}, which expand and acquire masses via violent merging processes. When two or more galaxy clusters collide, they generate internal merger shocks on a cosmological scale, and most of the released kinetic energy is dissipated in the surrounding intra-cluster medium (ICM) in the form of thermal energy \citep{2002ASSL..272....1S,2003PhPl...10.1992S}. Investigation of merger shocks gives important information on the cluster dynamics, allowing us to study the thermal history of galaxy clusters \citep{2001ApJ...563...95M,2005ApJ...627..733M,2007PhR...443....1M}. Hence observation of merging clusters is the best way to study, simultaneously, cosmic-ray acceleration over $\sim$ 1 Mpc cluster volume; interaction of these relativistic particles with cluster magnetic fields; the self-interaction properties of dark matter; and the role of shocks in the ICM.     

\par {Extended diffuse radio sources, linked with galaxy clusters, are synchrotron radiation characterised by low surface brightness ($\mu$Jy/arcsec$^{2}$) and steep spectrum ($\alpha$ $\leq$ -1; $S_{\nu}$ $\propto$ $\nu^{\alpha}$, where $\alpha$ is the spectral index and $S_{\nu}$ is the flux density at frequency $\nu$)}. The diffuse giant cluster-wide radio sources are currently grouped mainly in two classes: radio halos and relics \citep[e.g.][]{2019SSRv..215...16V}. The radio halos are located near the core of a cluster, with typical size $\gtrsim$ 1 Mpc. The relic sources are similar to the halos in their low surface brightness and large size, but are found typically in peripheral regions of the cluster within their virial radius. Radio relics are thought to trace radio-emitting particles accelerated at the location of { merger shocks} \citep{2004MNRAS.347..389H,2009A&A...506.1083V,2011A&A...528A..38V}. These shocks, expanding with high velocity (Mach number $\mathcal{M}$ $\sim$ 1-3), can accelerate electrons to high energies and compress magnetic fields, giving rise to synchrotron radiation being emitted from large regions. The accelerated particles will have a power-law energy distribution, as well as magnetic fields aligned parallel to the shock-front. This is in agreement with their elongated structure almost perpendicular to the merger axis. In some of the clusters, there is good agreement found between the position of X-ray shock-fronts and radio relics \citep[e.g.][]{2019SSRv..215...16V}. The radio properties of halos and relics are coupled to the global properties of their host galaxy cluster. {Observations suggest that clusters with the highest X-ray mass (and luminosity) have a higher probability of hosting radio halos or relics, or both. Furthermore, the monochromatic radio power of a halo and relic at 1.4 GHz correlates with the X-ray luminosity, mass, and temperature of a cluster \citep{2002ASSL..272..197G}. Recently, thanks to the improved sensitivities of radio telescopes and imaging techniques, radio observations have found that these diffuse radio sources could be associated with low massive clusters \citep{2018A&A...609A..61C}. }

\par { Moreover, these diffuse radio sources are mostly found in non-relaxed or merging galaxy clusters \citep{2010ApJ...721L..82C, 2015A&A...575A.127P}. They are, however, not detected in every merging cluster. This indicates that cluster merger shock waves (in the case of relics) or turbulence (in the case of halos) are not adequate to understand the radio emission mechanism behind diffuse radio sources. It is equally important to pay attention to the availability of magnetic fields, and the pool of pre-accelerated cosmic particles at cluster outskirts (for relics) and cluster-wide volume (for halos). {Theoretical models suggest that, when electrons are accelerated straight from the thermal pool, the required acceleration efficiency is high and it would be difficult to explain the observed radio luminosity of radio relics (\citealp{2011ApJ...734...18K}; \citealp[e.g.][]{2014IJMPD..2330007B}; \citealp{2016MNRAS.460L..84B}; \citealp{2020A&A...634A..64B})}. {For example, $\sim$ 10\% or more of shock acceleration efficiencies are expected to match the observed radio luminosities of relics} \citep{2020A&A...634A..64B}. Hence, alternatively, it is required that the electrons re-accelerated from the mild relativistic seed population. This assumption would lower the electron acceleration efficiency.} It is believed that the primary sources of these pre-accelerated (seed) particles are Active Galactic Nuclei (AGN) of radio galaxies and supernovae. In this scenario, it is possible that the electrons are ejected from the activities of radio galaxies (and/or supernovae) and diffused into the ICM. The cluster merger process re-energises these pre-existing electrons to travel across the $\sim$ Mpc distance. 

\par {A2384} is a nearby ($z$ = 0.092), low mass, and complex bimodal (having a two component - A2384(N) and A2384(S)) merging cluster \citep{2011A&A...525A..79M, 2014A&A...570A..40P,2019MNRAS.tmp.2668P}. A2384 is a peculiar cluster system that displays a dense X-ray filament ($\sim$ 700 kpc in length) between A2384(N) and A2384(S) (Fig.~\ref{Xray_radio_img}(a)). The optical data analysis of the galaxy distribution in the A2384 system, at all magnitude limits, also shows a very elongated structure along the north-south axis. The basic properties of the cluster are listed in Table \ref{sample}. A total of 56 cluster member galaxies have been identified, and both the X-ray gas and galaxy distributions approximately define the same bimodal structure \citep{2011A&A...525A..79M}. In a previous work \citep{2019MNRAS.tmp.2668P}, it was observed that, in the {\it Chandra} X-ray map, there is a bend visible in the X-ray bridge $\sim$ 800~kpc from the X-ray peak of A2384(N). We then obtained its GMRT 325 MHz data to study the radio environment around the cluster. We found a Fanaroff-Riley Class I (FR I henceforth) \citep{1974MNRAS.167P..31F} radio galaxy, a member of this cluster whose northern radio lobe pushes and distorts the X-ray filament eastwards \citep{2019MNRAS.tmp.2668P}. Such a displacement in the X-ray filament by the radio lobe is observed here for the first time on such a large scale. 
{We also found a surface brightness discontinuity and temperature jump in the direction of the north radio lobe of the FR I radio galaxy in the X-ray filament. This may suggest the presence of the shock corresponding to the Mach numbers $\mathcal{M}_{\delta}$ = 1.09 $\pm$ 0.06 and $\mathcal{M}_{\mathrm{T}}$ = 1.25 $\pm$ 0.035, respectively, {where $\mathcal{M}_{\delta}$ corresponds to the density jump and $\mathcal{M}_{\mathrm{T}}$ corresponds to the temperature jump across the shock}. This weak shock could be a result of the lobe-ICM interaction.} Further, morphological analysis of A2384 suggests that it is a post-merger cluster. The low mass cluster A2384(S) has passed through the more massive cluster A2384(N) and probably stripped it of a large amount of hot gas (and a large number of galaxies) in the direction of the north-south merger. In the XMM-Newton data, temperature and entropy are very high in the X-ray filament region, indicating the active dynamical region.


\par  In this paper, we show the radio observation of the A2384 galaxy cluster with the MeerKAT radio telescope.  This paper is organised as follows: Section~2 gives details of the MeerKAT radio observation of A2384; Section~3 describes the radio data reduction procedures; Section~4 presents our images and results; Section~5 gives the discussion; and finally, Section~6 provides the conclusions. In this paper, we have assumed $H_{0}$ = 70 km\,s$^{-1}$\,Mpc$^{-1}$, $\Omega_{\mathrm{M}}$ = 0.3 and $\Omega_{\Lambda}$ = 0.7. At redshift $z$ of 0.0943, 1$''$ = 1.75 kpc, and luminosity distance $D_{\mathrm{L}}$ = 432 Mpc. 

\begin{table*}
\centering
\caption{Properties of A2384. $M_{500}$ and $L_{500}$ taken from the MCXC catalogue \citep{2011A&A...534A.109P}. $M_{500}$ is the mass enclosed by a sphere within which the mean density is 500 time critical density of the Universe at $z$=0 and $L_{500}$ is [0.1--2.4] $keV$ band luminosities.}
\begin{tabular}{ccccccccccccccc}
\hline
\hline
Cluster & RA(J2000) & DEC(J2000) & $z$ & $M_{500}$ & $L_{500}$\\
    & h~~m~~s & d~~m~~s & & 10$^{14}$ $M_{\odot}$  &10$^{44}$ erg s$^{-1}$\\
\hline
A2384 & 21 52 21.9 & -19 32 48.6 &0.0943&  2.61 &1.66\\
\hline
\end{tabular}
\label{sample}
\end{table*}

\section{MeerKAT observation of A2384}
We observed A2384 with the MeerKAT telescope \citep{2009IEEEP..97.1522J,2018ApJ...856..180C} on 18 May 2019 with the full array. The MeerKAT array consists of 64 antennas which are located in the Karoo semi-desert, South Africa, and operated at L-band. A2384 was observed for a total of four hours, including overheads. The data were recorded with 4096 channels of the total bandwidth of 856 MHz. The centre frequency of the observation was 1283 MHz. The maximum and minimum MeerKAT baselines are 8 km and 29 m, respectively. This allows observation of any extended structure up to the size of $\sim$ 27$'$. In our observation, the primary flux calibrator was J1939-6342 (RA:19h39m25.05s, Dec:-63$^{\circ}$42$'$43.6$''$), and secondary gain calibrator was J2206-1835 (RA:22h06m10.33s, Dec:-18$^{\circ}$35$'$39$''$). The gain calibrator was observed for two minutes after every 15 minute target scan. The MeerKAT data were converted from the original MVFv4 format to a CASA Measurement Set (MS) using the KAT Data Access Library (KATDAL) package\footnote{https://github.com/ska-sa/katdal}. Subsequent initial flagging, cross-calibration, second-round flagging, imaging and self-calibration were performed using two different approaches - (1) the \textbf{C}ontainerised \textbf{A}utomated \textbf{R}adio Astronomy \textbf{Cal}ibration (CARACal) pipeline \citep{2020arXiv200602955J}, and (2) the manual \textbf{C}ommon \textbf{A}stronomy \textbf{S}oftware \textbf{A}pplication (CASA) \citep{2007ASPC..376..127M} data analysis. The reason for using these two methods is to compare the outcome of the MeerKAT data with two independent data analysis techniques. Below we have outlined our two data reduction procedures.    
\section{MeerKAT data reduction}
 \subsection{CARACal pipeline}
\par CARACal\footnote{Formerly known as MeerKATHI} is a pipeline being developed by an international collaboration, primarily between the South African Radio Astronomy Observatory, INAF - Osservatorio Astronomico di Cagliari and Rhodes University. CARACal has been designed to reduce continuum and spectral line data of MeerKAT and other radio interferometric telescopes, including JVLA and GMRT. CARACal is based on the Stimela\footnote{https://github.com/ratt-ru/Stimela} radio interferometry pipelining framework, which is platform-independent through the usage of containerisation technologies\footnote{https://www.docker.com/resources/what-container} (Docker, Singularity, Podman, and uDocker), and which allows seamless usage of different software suites, making it ideal for the creation of pipelines.
\par CARACal uses a collection of publicly available data reduction software as well as software developed in-house by the pipeline developers. A thorough description of the CARACal pipeline will be given in Makhathini et al. (in preparation). Here we describe the pipeline usage briefly. The CARACal pipeline is controlled via a configuration file which allows for user input in the data reduction process. Once run, the pipeline requires no further input from the user. The data products, including reduced visibility data, continuum and spectral images (the latter incorporating moment maps) as well as diagnostic plots, are provided by the pipeline. Below we describe the reduction of our MeerKAT data through the CARACal pipeline.

\par The standard configuration of the CARACal pipeline includes flagging, cross-calibration and self-calibration processes. For our MeerKAT dataset, the known radio frequency interference (RFI) channels are 856 to 880 MHz, 1419.8 to 1421.3 MHz, and 1658 to 1800 MHz, all of which are flagged. Calibrator and target fields were flagged through CARACal by using AOFlagger \citep{offringa-2012-morph-rfi-algorithm} to automatically excise RFI. To do this, we used custom strategies (calibrator- and target-data specific) observed to work best with MeerKAT data, which we utilised in this reduction. The cross-calibration scheme used by CARACal at the time of our reduction was fairly standard, including setting the flux scale and deriving corrections for residual delay calibration, bandpass and time-varying gain. We used the \cite{ReyATNF.352..508R} scale to fix the absolute scale of the flux calibrator. 

\par The CARACal pipeline uses standard tasks from the CASA suite for cross-calibration, and the Cubical software \citep{2018MNRAS.478.2399K} for self-calibration. After applying all the corrections to the target data, we channel-averaged the dataset by a factor of five channels while splitting (consistent with our science aims, since the source we target is more or less in the central part of our field, reducing the effect of smearing through the channel averaging). To deconvolve and image the target data, the WSClean imager \citep{2014MNRAS.444..606O} was used, with the multiscale and wideband deconvolution algorithms enabled to facilitate imaging diffuse emission expected for our field. Deconvolution was jointly performed in five sub-band images: 856-1026 MHz (163 total channels with central frequency 941 MHz), 1026-1198 MHz (164 total channels with central frequency 1112 MHz), 1198-1369 MHz (164 total channels with central frequency 1283 MHz), 1369-1540 MHz (164 total channels with central frequency 1454 MHz), and 1540-1712 MHz (164 total channels with central frequency 1626 MHz). In a joined-channel deconvolution mode of WSClean, it generates the multi-frequency synthesis (MFS) map; and in the present MeerKAT data, the central frequency of the MFS map is 1283 MHz, which is a full bandwidth map. In WSClean, each of the sub-bands is deconvolved separately with an initially high mask of 40$\sigma_{\mathrm{rms}}$ (using the auto masking function provided by WSClean), to generate an artefact-free model of the target field for the self-calibration process. This masking threshold is iteratively reduced to a value of 3$\sigma_{\mathrm{rms}}$, the final iteration of imaging. We made images with different Briggs robust weighting schemes in order to reduce the sidelobes of the point-spread function. To image the extended sources in the data, we used robust weighting of +1, which is close to the natural weighting, giving more weight to each of the inner baselines. This resulted in an angular resolution of 30$''$ $\times$ 22$''$ of sub-band 1 to 19$''$ $\times$ 13$''$ of sub-band 5. To match the resolution of every band image, we used the \texttt{beamshape} parameter of WSClean and fixed the lower resolution of 30$''$ $\times$ 22$''$ for all sub-band images.  We performed several rounds of self-calibration until the rms noise level had reduced from the previous round. 

\par In our analysis, we found that robust +1 weighting gives the highest sensitivity to the extended emission. We also compared the detected extended features in the robust +1 image with the image generated using the combination of robust 0 weighting and $uv$ tapering of 25$''$. In both images, there is good agreement between source locations and sizes. In order to detect the blended and unresolved point sources, we also generated a high-resolution uniform image of beam size 4$''$ $\times$ 4$''$. 

\subsection{Manual data reduction}
We analysed the MeerKAT data using the CASA 5.1.0 version. Firstly, we removed the channels with known RFI from the data, as mentioned above. Secondly, we used CASA-based automated RFI removal tasks, \texttt{rflag} and \texttt{tfcrop}. This automated flagging was applied to both calibrators and target sources. The primary calibrator (J1939-6342) was used to derive antenna delay and bandpass corrections, as well as to fix the absolute flux scale. Antenna-based time-dependent complex gain was derived using the secondary calibrator (J2206-1835), the solutions of which were then applied to all the other fields. After calibration, we again ran \texttt{rflag} and \texttt{tfcrop} on the calibrators and target fields for more precise RFI removal from the calibrated visibilities. Thirdly, we repeated the cross-calibration steps; and finally, we split the target source averaged over five channels. In the deconvolution process, we used wide-field, wideband imaging algorithms such as \texttt{w-terms} and \texttt{mt-mfs} (with Taylor term = 2) \citep{2011A&A...532A..71R}. We performed several rounds of self-calibration until the rms remained consistent. After each of the self-calibration rounds, we checked and flagged any bad data remaining in the visibilities using \texttt{rflag} and \texttt{tfcrop}. We made final images of the A2384 field, using natural weighting to map extended emission. 

\subsection{Comparison between the CARACal pipeline data reduction and Manual data reduction}
Final images produced by these two methods were compared. There was agreement between the two sets of images. We found, however, that the image(s) generated by the CARACal pipeline had fewer artefacts and higher signal to noise (SNR) compared with the manually produced image. For the rest of the analysis, therefore, we used images produced only by the CARACal pipeline.  

\section{Results and images}
\par
\begin{center}
\begin{figure*}
    \centering
    \begin{subfigure}[t]{0.45\textwidth}
        \includegraphics[width=1\textwidth]{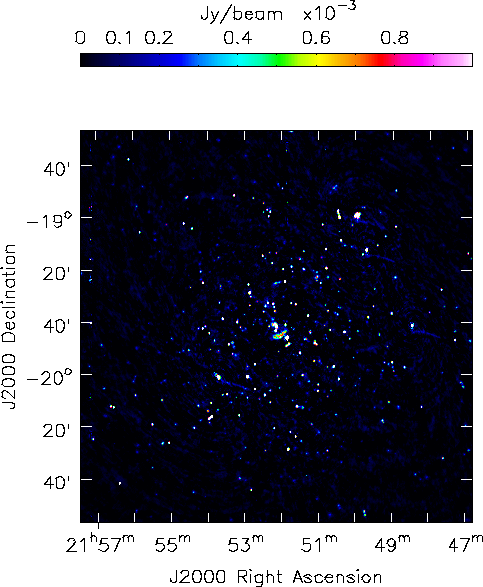}
        \caption{}
        \label{rfidtest_xaxis1}
    \end{subfigure}
    \begin{subfigure}[t]{0.45\textwidth}
        \includegraphics[width=1\textwidth]{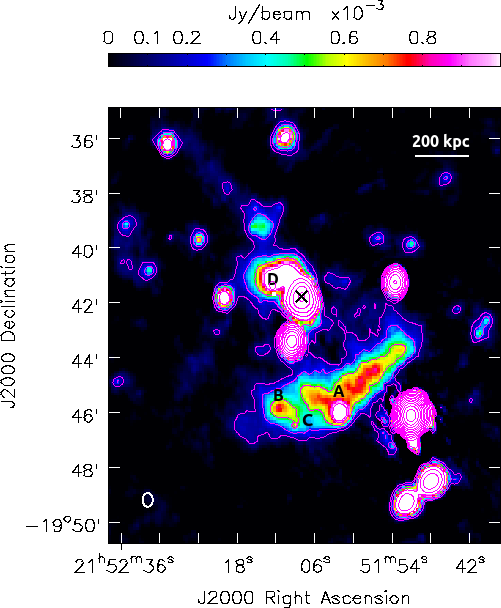}
        \caption{}
        \label{rfidtest_yaxis2}
        \end{subfigure}
       \caption{(a) MeerKAT image of 1$^{\circ}$ surrounding region of A2384.  (b) Central zoom region of A2384. We also drawn contours (magenta color) on it. The contour levels start at 3$\sigma$ and increase by factors of 2. Negative contours are drawn with a dashed line. The beam size and shape are shown bottom-left. We marked (A,B,C) discrete and unresolved radio sources in the extended relic region, and (D) in the lobe region. We marked the position of the FR I radio galaxy with an `$\times$'. In the MeerKAT radio images, the beam size is 30$''$ $\times$ 22$''$ and 1$\sigma$ = 44 $\mu$Jy beam$^{-1}$. }
    \label{MeerKAT_radio_img}        
\end{figure*}
\end{center}

\begin{figure*}[!ht]
    \centering 
  \includegraphics[width=1\textwidth]{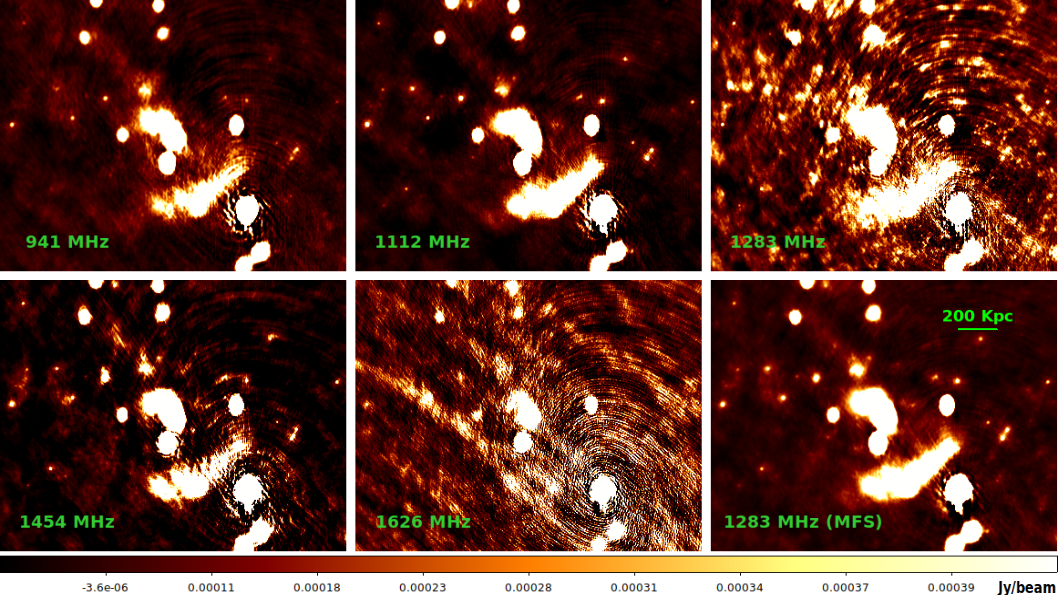}

\caption{Five sub-bands and MFS images of MeerKAT full bandwidth of 820 MHz. Each sub-band image has 164 MHz wide bandwidth. In each of the sub-band images, we have given the central frequency.}
\label{fig:sub_band_img}
\end{figure*}

 Our MeerKAT observation revealed many new features in the A2384 complex merging cluster. Some of the radio features had been missing from the previous 325 MHz GMRT image. We show the full MeerKAT radio image of the A2384 and surrounding regions in Fig.~\ref{MeerKAT_radio_img}(a) for the 1$^{\circ}$ area. In Fig.~\ref{fig:sub_band_img}, we give each sub-band image around the central region of A2384. In \cite{2019MNRAS.tmp.2668P}, we reported a peculiar detection of the FR I radio galaxy LEDA 851827 (J2000 RA:21h52m08.1s; Dec:-19$^{\circ}$41$'$27$''$) associated with A2384(S). Its northern radio lobe distorts the X-ray filament in the easterly direction. In the MeerKAT data, we also noticed the northern radio lobe of 208 $\times$ 270 kpc$^{2}$ (119$''$ $\times$ 154$''$), but that southern radio lobe was missing (Fig.~\ref{MeerKAT_radio_img}(b)). In the MeerKAT high-resolution image, we also detected an unresolved point source embedded in the lobe at position RA: 21h52m12.45s, Dec:-019$^{\circ}$41$'$22.15$''$.
\subsection{Discovery of a radio relic}
In our MeerKAT images, we discovered an extended radio source at the bottom of the A2384(S) cluster (Fig.~\ref{Xray_radio_img}(b)). This extended radio source is situated perpendicular to the A2384 merger axis. The physical size of the radio source is $\sim$ 824 ({\it l}) $\times$ 264 ({\it w}) kpc$^{2}$ (or angular size $\sim$ 471$''$ $\times$ 151$''$).  The geometry, location and size of this radio source suggest that it is a radio relic associated with merger shock and cluster A2384. The radio relic is extended from the south-east to the north-west. The south-eastern part is close to the A2384(S). The shock-front of the relic is located at a distance of $\sim$ 3.7$'$ from A2384(S) and 14.6$'$ from A2384(N). In the high-resolution image, three discrete and unresolved point sources were detected in the relic. We have identified these point sources in Fig.~\ref{MeerKAT_radio_img}(b).  In Fig.~\ref{Xray_radio_img}(c), we plot the GMRT and MeerKAT radio contours together. In the GMRT data, the radio relic was not detected clearly, and misinterpreted as the southern lobe of the FR I radio galaxy. We discuss this further in section \ref{relic-lobe}. We were able to see this radio relic to its full extent in three sub-band maps (941, 1112, and 1454 MHz) only, while, due to severe RFI in the 1283 MHz and 1626 MHz bands, we could not observe the whole relic.

\subsection{Diffuse radio source in the X-ray bridge}
{\par In the MeerKAT image, we noticed a new faint and diffuse radio source located at the north end of the northern lobe of the FR I radio galaxy, close to the X-ray filament distortion. This diffuse radio source is $\sim$ 3.1$'$ from the FR I core galaxy in the northerly direction. We were unable to find any optical or radio counterpart to this diffuse radio emission in high-resolution (4$''$ $\times$ 4$''$) MeerKAT map. This suggests that this potential radio source is not related to any galaxy, but is instead a diffuse radio emission known as a radio ridge associated with the A2384 cluster dynamical process. The morphology of this radio source is roughly circular as compared with that of mini radio halos found in the core of relaxed galaxy clusters. The size of this radio ridge is $\sim$ 182 $\times$ 129 kpc$^{2}$ (104$''$ $\times$ 74$''$). This radio ridge can be seen to full extent only in the 941 MHz frequency sub-band image as being similar to the MFS image. The present MeerKAT data are not enough to further study of this candidate radio ridge except its location and size.}  We have reported the integrated flux density values of radio relic, lobe, candidate radio ridge and discrete point sources in Table \ref{flux_val}. In this paper, all integrated flux density values are reported to have above 3$\sigma$ background rms noise.

\subsection{Spectral index values and maps}
\par
\begin{center}
\begin{figure*}
    \centering
    \begin{subfigure}[t]{0.45\textwidth}
        \includegraphics[width=1\textwidth]{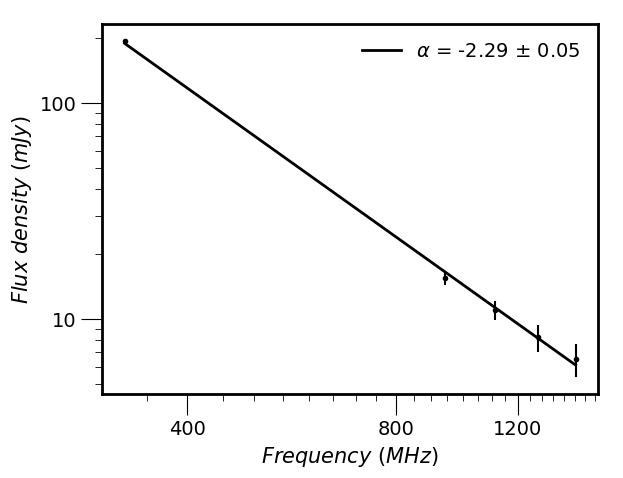}
        \caption{}
        \label{rfidtest_xaxis1}
    \end{subfigure}
    \begin{subfigure}[t]{0.45\textwidth}
        \includegraphics[width=1\textwidth]{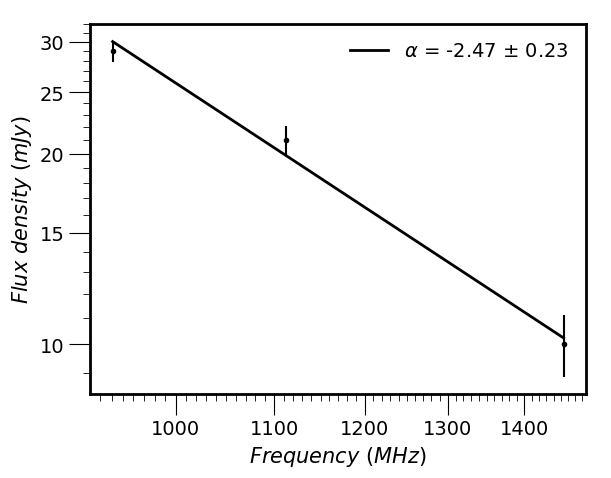}
        \caption{}
        \label{rfidtest_yaxis2}
        \end{subfigure}
    \caption{(a) Spectral index plot of the radio lobe, and (b) Spectral index plot of the radio relic}
    \label{radio_spectra}        
\end{figure*}
\end{center}

 We calculated the spectral index values for the radio relic, northern radio lobe and candidate radio ridge, and those of the relic and lobe are plotted in Fig.~\ref{radio_spectra}. Before calculating the spectral index values of the relic and lobe, we subtracted the flux densities of discrete point sources (blended with relic and lobe) from the total integrated flux densities of the relic and lobe, respectively. As mentioned above, the current MeerKAT data are not sufficient to image the relic and radio ridge at all sub-band images due to severe RFI. We calculated the relic spectral index to be between 941 MHz, 1112 MHz and 1454 MHz frequencies. We measured the north lobe spectral index value between the 325 MHz, 941 MHz, 1112 MHz, 1283 MHz and 1454 MHz frequencies. To match the beam sizes, we convolved the GMRT 325 MHz image to the closest 30$''$ $\times$ 25$''$ beam. The GMRT and MeerKAT flux densities are on the \cite{2012MNRAS.423L..30S} and \cite{ReyATNF.352..508R} scales, respectively. We do not know the agreement between these two scales, as each is derived for a different hemisphere and calibrator. Hence it is not possible to match the calibration between GMRT and MeerKAT. We found the radio relic spectral index to be $\alpha$ = -2.5 $\pm$ 0.23 and the radio lobe spectral index to be $\alpha$ = -2.3 $\pm$ 0.05. We also found that, for the candidate radio ridge, the spectral index value for 941 MHz and MFS (1283 MHz) data was $\alpha$ $\sim$ -1.88.

\par We used the BRATS software \citep{2013MNRAS.435.3353H,2016MNRAS.458.4443H} to generate the spectral index map of A2384 (radio relic and lobe). In order to produce the spectral index map, we used images at only two frequencies (941 MHz and 1454 MHz), which have the highest SNR when compared with other frequency maps. We masked the blended unresolved point sources in the relic and lobe emission. We then convolved these maps to 32$''$ $\times$ 32$''$ beam size to smooth the images, and generated the spectral index map, as shown in Fig.~\ref{spx_map}(a) and its error map in Fig.~\ref{spx_map}(b). 

\par In order to calculate the spectral index values, BRATS software uses a weighted least squares method (pixel by pixel), where the weights are calculated by $\omega$ = 1/$\sigma^{2}$ and $\sigma$ is the error on given flux measurements. We set the \texttt{sigma} parameter to 3 so that, in the spectral index calculation, the BRATS software would include only those pixels whose flux is $>$ 3$\sigma_{\mathrm{rms}}$. We used the remaining standard default parameters in the BRATS to generate the spectral index and error maps.

\subsection{Radio power at 1.4 GHz measurement}
\par For the A2384 relic, lobe and candidate radio ridge, we estimated its K-corrected (rest-frame) radio power ($P_{1.4\mathrm{GHz}}$) using the following: 
\begin{equation}
P_{1.4\mathrm{GHz}} = \frac{4 \pi D_{\mathrm{L}}^{2}S_{v}}{(1+z)^{1+\alpha}},
\end{equation}
\par where $D_{\mathrm{L}}$ is the source luminosity distance at $z$ = 0.094,  and $S_{v}$ is the source flux density in the MFS image. For the radio relic, $\alpha$ = -2.5 gives radio power at 1.4 GHz to be $P_{1.4\mathrm{GHz}}$ = (3.87 $\pm$ 0.40) $\times$ 10$^{23}$ W Hz$^{-1}$. For the radio lobe, $\alpha$ = -2.3 gives radio power to be $P_{1.4\mathrm{GHz}}$ = (2.05 $\pm$ 0.21) $\times$ 10$^{23}$ W Hz$^{-1}$. For the candidate radio ridge, $P_{1.4\mathrm{GHz}}$ = (4.60 $\pm$ 0.46) $\times$ 10$^{22}$ W Hz$^{-1}$.

\par {The procedure we followed to estimate the error in the flux density measurements is outlined below}. There are two primary sources of errors in the flux density measurements: (1) an error due to the uncertainties in the flux densities of the unresolved source(s) used for calibration of the data. For the MeerKAT data, we assumed this error to be $\sim$ 10$\%$ (private communication with MeerKAT commissioning team); and (2) since the diffuse radio sources are extended, the errors in their flux density estimations will be the rms in the image, multiplied by the square root of the ratio of the solid angle of the source to that of the synthesised beam, which is the number of beams across the source. Here we note that our observation was affected by RFI so our flux density error could have been underestimated. These two sources of errors are unrelated, so they are added in quadrature to estimate the final error on the flux densities of the extended sources, as shown below:
\begin{equation}
\Delta{S} = [(\sigma_{\mathrm{amp}}S)^2 + (\sigma_{\mathrm{rms}}\sqrt{n_{\mathrm{beams}}})^2)]^{1/2},
\label{cal_err}
\end{equation}
  \par where $S$ is the flux density, $\sigma_{\mathrm{amp}}$ is the flux calibration uncertainty (10\%), $\sigma_{\mathrm{rms}}$ is the image rms noise, and $n_{\mathrm{beams}}$ is the number of beams in the full extent of the source.
  
\begin{center}
\begin{figure*}
    \centering
    \begin{subfigure}[t]{0.47\textwidth}
        \includegraphics[width=1\textwidth]{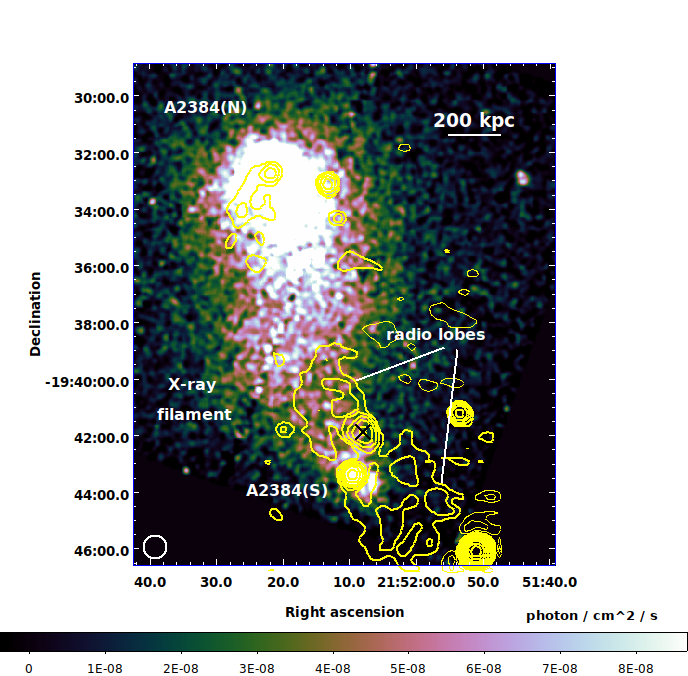}
        \caption{}
        \label{rfidtest_xaxis1}
    \end{subfigure}
    \begin{subfigure}[t]{0.48\textwidth}
        \includegraphics[width=1\textwidth]{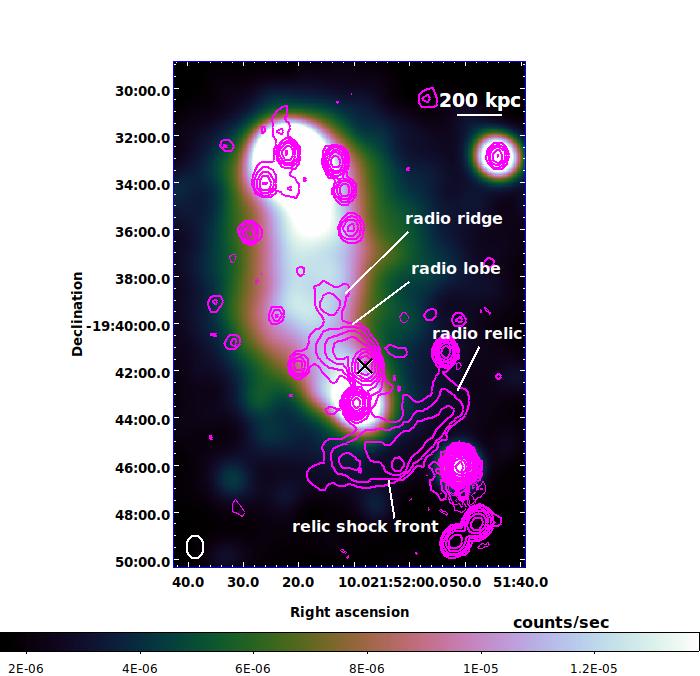}
        \caption{}
        \label{rfidtest_yaxis2}
        \end{subfigure}
     \begin{subfigure}[t]{0.44\textwidth}
        \includegraphics[width=1\textwidth]{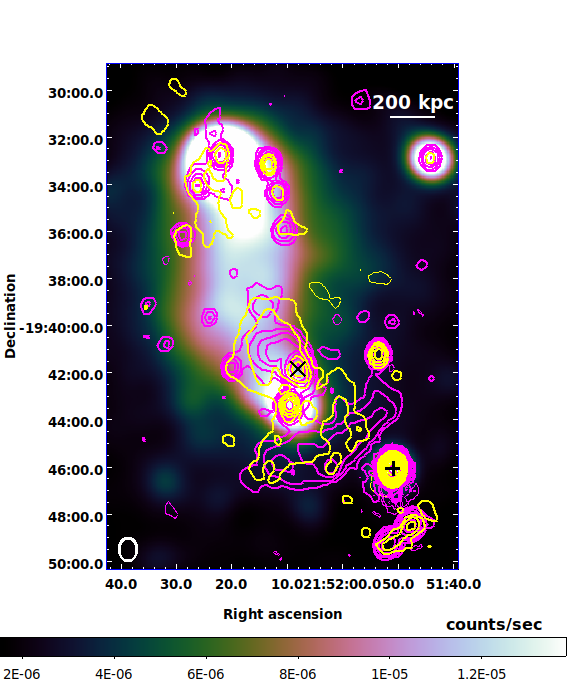}
        \caption{}
        \label{rfidtest_xaxis1}
    \end{subfigure}
      
    \caption{(a) A2384 {\it Chandra} X-ray color image. The yellow contours show the GMRT 325 MHz observation. (b) MeerKAT radio contours (magenta) on A2384 XMM-{\it Newton} image. In this image, we marked three extended radio sources. (c) MeerKAT (magenta) and GMRT (yellow) radio contours on A2384 XMM-{\it Newton} image. In all images, contour levels start at 3$\sigma$ and increase by factors of 2. Negative contours are drawn with a dashed line. The beam size and shape are shown at the bottom-left of all images. In the GMRT radio image (a), the beam size is 25$''$ $\times$ 25$''$ and 1$\sigma$ = 0.8 mJy beam$^{-1}$. For the MeerKAT radio image (b), the beam size is 30$''$ $\times$ 22$''$ and 1$\sigma$ = 44 $\mu$Jy beam$^{-1}$. For the GMRT and MeerKAT images in (c), the beam sizes are 30$''$ $\times$ 25$''$ and 30$''$ $\times$ 22$''$, respectively. For the GMRT, 1$\sigma$ = 1.4 mJy beam$^{-1}$, and for the MeerKAT rms is same as (b). We marked the position of the FR I radio galaxy (LEDA 851827) with a black `$\times$' in all images. In (c) we also marked the position of the bright radio point source (PKS 2149-20, J2000 RA:21h51m51.03s, Dec:-19$^{\circ}$46$'$05.51$''$) with a black `+'.}
    \label{Xray_radio_img}        
\end{figure*}
\end{center}

\begin{table*}
\centering
\caption{Flux density measurements from GMRT and MeerKAT sub-bands radio observations of A2384. The 325 MHz flux densities are from GMRT. 'MFS' flux densities correspond to the full MeerKAT observations between 941 MHz to 1626 MHz at the centre frequency of 1283 MHz. }

\begin{tabular}{cccccccc}
\hline
Frequency (MHz) & 325 & 941  & 1112 & 1283 & 1454 & 1626 & MFS\\ 
\hline
RMS (mJy beam$^{-1}$) &1.5 & 0.12 & 0.072 & 0.14 & 0.047 & 0.2 & 0.044\\
\hline
Radio relic (mJy) &- &29$\pm$3.1 & 21$\pm$2.2 & - & 10$\pm$1.1& - & 18.8$\pm$1.9\\
Radio lobe  (mJy) &193$\pm$21.5 &15.4$\pm$1.6 & 11$\pm$1.1 &8.2$\pm$1.0 & 6.5$\pm$0.7& - &10$\pm$ 1.0\\
Radio ridge (mJy) & -& 4$\pm$0.5 & -& -& -& -&2.23$\pm$0.2\\
A (mJy)  & - &7.33$\pm$0.74 & 5.73$\pm$0.57 & 4.28$\pm$0.45&3.25$\pm$0.32 & - &4.66$\pm$0.46\\
B (mJy)  & -&5.44$\pm$0.55  &  6.39$\pm$0.64  &  2.82$\pm$0.31  &  3.16$\pm$0.32&- &4.31$\pm$0.43\\
C (mJy) & - &0.66$\pm$0.13 &  0.45$\pm$0.08  &  0.13$\pm$ 0.14&  0.23$\pm$0.05 &- &0.33$\pm$0.05\\
D (mJy) & 6.5$\pm$1.6 & 2.6$\pm$0.28 & 2.0$\pm$0.21 & 1.6$\pm$0.21 & 1.0$\pm$0.10 &- & 1.6$\pm$0.16\\
\hline
\label{flux_val}
\end{tabular}

\end{table*}

\section{Discussion}
\subsection{Radio shock and Mach number}
\par Extended diffuse radio relic sources are tracers of shock waves in cluster merger events \citep{1998A&A...332..395E,2001A&A...366...26E}. Numerical simulations predict that a wide range of shocks pass through the ICM and intergalactic medium (IGM), and affect the formation of the group of clusters in large scale structure evolution \citep{2000ApJ...542..608M,2003ApJ...593..599R,2006MNRAS.367..113P,2007MNRAS.375...77H}. In astrophysical observations, there are usually two types of shock that exist: `internal' and `external', both of which control the growth of the large scale structure. The internal (or merger) shocks of velocities $\geq$ 1000 km\,s$^{-1}$ impact inner cluster material, which is already heated by the gravitational potential and supplies extra heating almost equivalent to the cluster temperature ($\sim$ 10$^{7}$ $K$). {In addition to this, a few per cent of the kinetic shock energy is converted into the (re-)acceleration of relativistic particles through a first-order Fermi acceleration mechanism (Fermi-I), which is thought to generate a radio relic. By contrast, the external (or accretion) shocks act across the cluster's peripheral region and increase the temperature of the outer low dense and cold gas. Accretion shocks could be associated with filament radio sources. {Furthermore, structure formation shocks passing through filaments are predicted to be narrow, and resulting synchrotron radio emission is highly polarised and the spectrum is flat ($\alpha$ $>$ -1)} \citep{2008Sci...320..909R,2011ApJ...735...96S}}. 
\par In the case of the diffusive shock acceleration (DSA) mechanism \citep{1983RPPh...46..973D}, the radio Mach number is given by: 
\begin{equation}
\mathcal{M} = \left(\frac{2\alpha_{\mathrm{inj}} + 3}{2\alpha_{\mathrm{inj}} -1}\right)^{1/2},
\end{equation}
\par  where $\alpha_{\mathrm{inj}}$ is the injection index. The injection index is flatter than the integrated spectral index by 0.5 \citep{1962SvA.....6..317K}  

\begin{equation}
\alpha_{\mathrm{int}} = 0.5+\alpha_{\mathrm{inj}}.    
\end{equation}

{With the spectral index value of the A2384 relic, $\alpha_{\mathrm{int}}$ = -2.5, we achieved $\mathcal{M}$ $\sim$ 1.41 $\pm$ 0.07. The relic has a very steep spectrum, and it is very challenging to describe the acceleration of the particles from the thermal pool, because it requires high acceleration efficiency. Alternatively, another theoretical model could suggest that the merger shocks re-accelerate a pre-existing population of relativistic particles. Below we show the possible connection of the relic with the radio galaxy which can supply the ultra-relativistic seed particles.}


\subsection{A2384 merger scenario and radio relic-lobe connection}
\label{relic-lobe}
\par In an earlier paper \citep{2019MNRAS.tmp.2668P}, we studied the X-ray morphology of the A2384 galaxy cluster, the results of which suggest that A2384 is a post-merger cluster. \cite{2011A&A...525A..79M} have also argued that A2384 is a post-merger cluster of two unequal mass clusters. During the interaction of the clusters, sub-cluster A2384(S) has passed through A2384(N) and is likely to have removed a large amount of hot gas (and a number of galaxies) from both systems in the direction of the merger. The A2384(S) would then be the front of the merging system. The velocity of the BCG of A2384(S) is $\sim$ 28,696 km\,s$^{-1}$ (private communication with Maurogordato), which is approximately 1000 km\,s$^{-1}$ higher than the BCG of A2384(N). This interaction of A2384 sub-clusters could lead the generation of shock waves which travel towards the merging direction. Detection of the relic supports the idea that A2384 is a post-merger system. 


\par Typically, powerful merger shocks produce shock waves with Mach number $\mathcal{M}$ in the range 2-3 \citep[e.g.][]{2019SSRv..215...16V}. These merger shocks, however, are not enough to accelerate the cosmic particles from the thermal pond to relativistic energies \citep{2011ApJ...734...18K, 2012ApJ...756...97K}. Observations indicate that extended radio sources (halos and relics) are the results of either turbulent re-acceleration (halos) in the volume of the clusters, or the propagating of merger shock waves (relics) at the outskirts of the clusters. This may suggest that a seed population of mildly relativistic or fossil electrons must already be present before the merger shocks propagate. {Cluster merger shocks can then re-trigger these pre-existing electrons. It is, however, important to investigate the origin and nature of the fossil electrons to understand their seed spectra. One possible explanation is that they originated from the AGN of the radio galaxies of clusters, and re-accelerated at the time of the cluster merger shocks.} Several recent studies have found the connection between tailed radio galaxies and relics \citep{2014ApJ...785....1B,2016MNRAS.460L..84B,2017NatAs...1E...5V,2018ApJ...865...24D}. In the case of A2384, we detected the diffuse radio sources (relic and candidate radio ridge) in the vicinity of the FR I radio galaxy which could be a natural source of the seed particles. There are discrete radio sources (probably AGNs) also blended with the relic, which could also be sources of the relativistic particles, but we do not have access to their redshifts information, so it would be difficult to prove their association with the relic. Given the current MeerKAT data, we are unable to see the southern radio lobe in the direction of the radio relic. 
 
\par It is surprising that we had not detected the extended radio relic in the previous GMRT data; we had confused the relic with the south lobe of the FR I radio galaxy (Fig.~\ref{Xray_radio_img}(a)). There are, however, several possible reasons for that. One of the main reasons is that there is a very bright and unresolved background point source (PKS 2149-20, $z$ = 0.424) situated 13$'$ away (RA: 21h51m50s; Dec:-19$^{\circ}$46$'$06$''$) from the phase centre and $\sim$ 2$'$ below the newly detected radio relic in the MeerKAT data (Fig.~\ref{Xray_radio_img}(c)). The integrated flux density of this point source is $S_{\mathrm{325MHz}}$ $\sim$ 4.4 Jy and $S_{\mathrm{1283MHz}}$ $\sim$ 2.1 Jy. Hence the quality of both the GMRT and MeerKAT images is degraded by the deconvolution errors. Both images are dynamic range limited towards this bright point source. Another reason could be that low-frequency observations are hampered by the strong RFI. This RFI is picked up mostly by the shorter baselines antennas (central square of the GMRT). These short spacing antennas are critical in the imaging of extended radio sources. Because of the excessive flagging ($\sim$ 50$\%$) of this data, however, it is difficult to image the relic in the GMRT data unambiguously. Likewise, in the MeerKAT observation, due to RFI, $\sim$ 40 \% of the data are not usable. Hence, the presence of the nearby bright point source, complex relic emission, and data loss due to the excessive RFI, could have prevented us from imaging the southern radio lobe of the FR I radio galaxy.

\subsection{Diffuse radio source in the hot X-ray filament}
{\par We detected a diffuse radio source located in the north of the northern radio lobe of the FR I radio galaxy, close to the A2384 filament distortion location. The radio source is small ($\sim$ 182 $\times$ 129 kpc$^{2}$), and could be considered as a candidate radio ridge generated in the X-ray hot and dense filament of A2384. The nature of the detected radio source is unclear, and difficult to study in detail. It is particular difficult to observe the non-thermal radio emission beyond the clusters within filaments. There are a handful of examples of diffuse radio emission found between two or more clusters \citep{2018MNRAS.478..885B,2019A&A...630A..77B,2019Sci...364..981G}.
This non-thermal radio emission could be a result of the (re-)accretion of matter on large scales filaments, which likely drives turbulence and shock waves on a different range of spatial scales. 


Recently, numerical simulation of a binary cluster collision \citep{2020PhRvL.124e1101B} has shown the radio ridges spreading across several Mpc distances between two galaxy clusters. This radio emission originates from the second-order Fermi acceleration (Fermi-II) of electrons interacting with turbulence. Due to this turbulence, magnetic fields are amplified, and electrons are scattered within it. Furthermore, weak shocks between the clusters may also compress the re-accelerated population of the electrons, as well as magnetic fields which eventually boost the radio signal. The source of turbulence could be the complex dynamics of substructures embedded in filaments between galaxy clusters. In the earlier X-ray analysis of A2384 \citep{2019MNRAS.tmp.2668P}, we found high temperature and entropy in the X-ray filament, suggesting an active dynamical region. We also found the presence of a weak shock ahead of the northern radio lobe. The puzzling diffuse radio source detected in the filament between A2384(N) and A2384(S) can be used to set the limits on the non-thermal energy and large scale magnetic fields in the X-ray filament. We were unable to image this candidate radio ridge to its full extent (compared with the MFS image), at all frequencies, with the exception of 941 MHz. The available MeerKAT data are not sufficient to understand the radio ridge formation mechanisms; neither are they sufficient to study the connection between radio lobe and ridge emission.}


\subsection{Equipartition magnetic field}
\par Assuming the equipartition magnetic field for the A2384 relic (which gives the same energy densities in magnetic fields and relativistic cosmic particles), then, using the following formula \citep{2004IJMPD..13.1549G}, we can estimate the magnetic fields in the A2384 relic: 
 
\begin{dmath}
u_{\mathrm{min}} = \xi(\alpha,v_{1},v_{2})\times(1+k)^{4/7}\times(v_{0})^{4\alpha/7}\times\\(1+z)^{(12+4\alpha)/7}\times(I_{0})^{4/7}\times(d)^{-4/7},\\
\end{dmath}

\par where $u_{\mathrm{min}}$ expressed in erg cm$^{-3}$; $\xi$ is the equipartition constant and function of $\alpha$ and frequency ranges $v_{1}$ and $v_{2}$; $\alpha$ is the injection index; $I_{0}$ (in mJy arcsec$^{-2}$) is the source brightness at the frequency $v_{0}$ (in MHz) and can be measured by dividing the integrated flux density of the relic by its solid angle (471$''$ $\times$ 151$''$); $z$ is the redshift of A2384; $k$ is the ratio of the energy in relativistic protons to that in electrons (= 1); and $d$ (given in kpc) is the depth of the relic ($d = \frac{824_{\mathrm{kpc}}+264_{\mathrm{kpc}}}{2}$). The equipartition magnetic field is then measured as at $v_{0}$ = 1283 MHz: 

\begin{equation}
B_{\mathrm{eq}} = \left(\frac{24\pi}{7} u_{\mathrm{min}}\right)^{1/2}.
\label{eqp}
\end{equation}

For frequency range 10 MHz to 100 GHz, we estimate that $B_{\mathrm{eq}}$ for the relic is $\sim$ 1.2 $\mu$G, and the northern radio lobe of the FR I radio galaxy is $\sim$ 1.5 $\mu$G.  

 {The revised formula for the equipartition magnetic field ($B'_{\mathrm{eq}}$) is derived using the electron energy by its Lorentz factor, $\gamma$, rather than using the frequency to calculate the magnetic field strength of synchrotron radio sources given by \citep[\& references therein]{2004IJMPD..13.1549G}: }

\begin{equation}
B'_{\mathrm{eq}} = 1.1\gamma_{\mathrm{min}}^{\frac{1-2\alpha}{3+\alpha}} B_{\mathrm{eq}}^{\frac{7}{2(3+\alpha)}},
\end{equation}

\par where $\gamma$ is the Lorentz factor (= 100) and $B_{\mathrm{eq}}$ is the value of the equipartition magnetic field obtained in equation \ref{eqp}. This gives the $B_{\mathrm{eq}}$ for the relic source as $\sim$ 5 $\mu$G, and the radio lobe as $\sim$ 5 $\mu$G.




\subsection{Spectral index map}


\par
\begin{center}
\begin{figure*}
    \centering
    \begin{subfigure}[t]{0.45\textwidth}
        \includegraphics[width=1\textwidth]{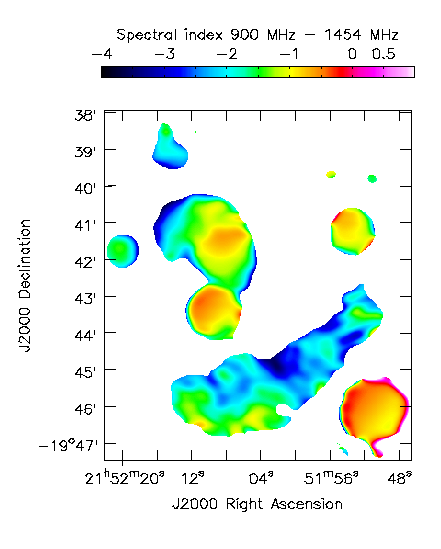}
        \caption{}
        \label{rfidtest_xaxis1}
    \end{subfigure}
    \begin{subfigure}[t]{0.45\textwidth}
        \includegraphics[width=1\textwidth]{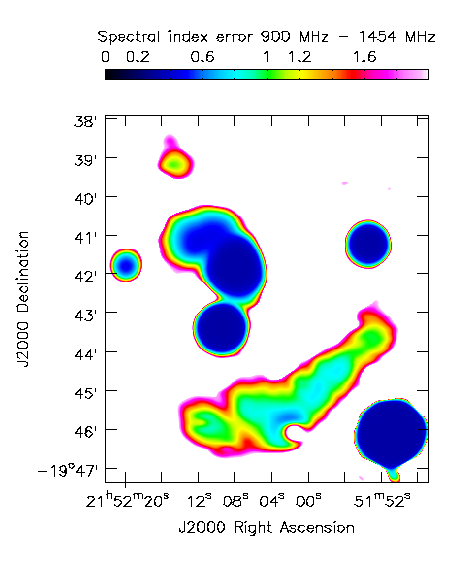}
        \caption{}
        \label{rfidtest_yaxis2}
        \end{subfigure}
    \caption{(a) Spectral index map of A2384 (relic and lobe). We generated this map between 941 MHz and 1454 MHz images. (b) Spectral index error map corresponding to (a).}
    \label{spx_map}        
\end{figure*}
\end{center}

 The spectral index map is given in Fig.~\ref{spx_map}(a) and the error map in Fig.~\ref{spx_map}(b). In the spectral index map, we found the average spectral index of the entire relic to be $\sim$ -2.0, and that of the north radio lobe to be $\sim$ -1.73. The spectral index distribution in the relic is patchy. 
Overall, the south-eastern part of the relic is flatter, which means that compared with the north-western part, it is a dynamically more active. Furthermore, there is a gradient visible in the south-eastern part of the relic. The spectral index of the outer edge (shock-front) of the relic is flat. According to the DSA model, at the location of the shock-front, particles are re-accelerated, and hence the spectrum is flatter. By contrast, in the downstream region, steeper radio emission is found. In the present case, the spectrum is steepening towards the cluster centre, which suggests the aged population of electrons and synchrotron losses is high. Hence our findings are similar to those of the DSA model. We note that the errors associated with the spectral index values are high, which are largely systematic due to the low SNR of the data.
Further, the presence of imaging artefacts could affect the spectral index map \citep[and reference therein]{2014ApJ...785....1B}. As mentioned above, the presence of a bright point source just below the north-western relic part could affect the spectral index variations, as well as being the reason behind the patchy spectral index map. 
\par There is a gradient visible in the spectral index distribution in the north lobe of the FR I galaxy. The spectrum becomes steeper the further it moves from the core. The core of the FR I galaxy is flatter, and the spectrum becomes steeper near to the edge of the lobe. This trend is observed typically in the FR I types of radio galaxies. 

\subsection{P1.4 vs Mass relation}
\begin{figure}
    \includegraphics[width=\linewidth]{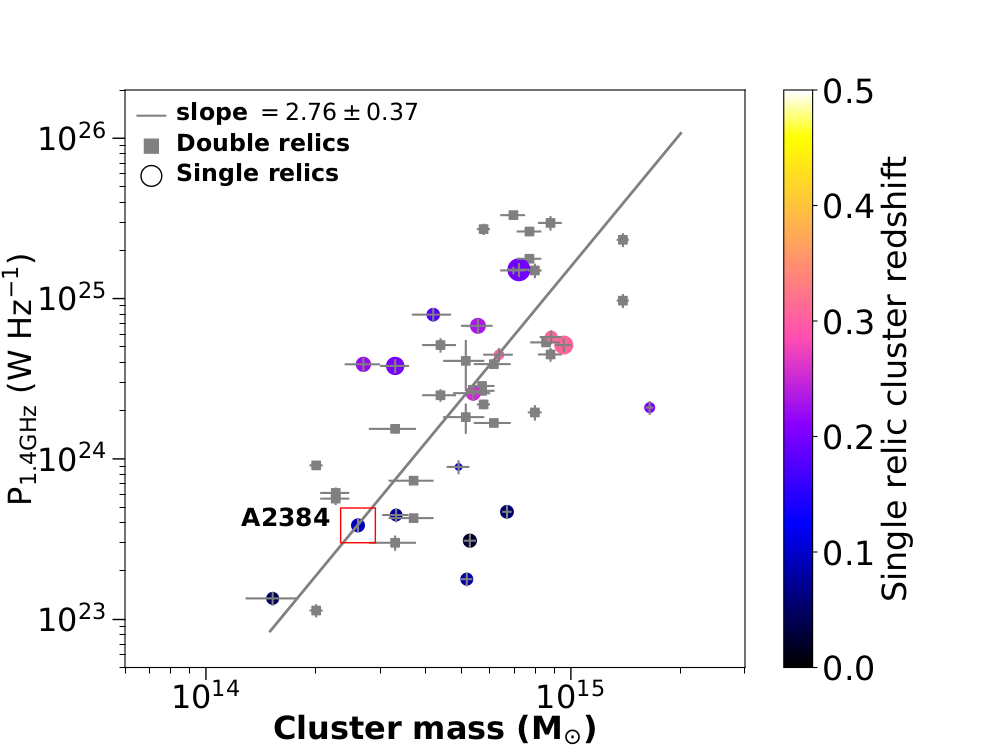}
    \caption{$P_{1.4}$ vs Mass of relic clusters. We have marked the position of A2384 with the red box. Each double relic clusters are plotted for their individual radio relic. For more details see \citet{2017MNRAS.472..940K}.}\label{p1.4vsmass}
\end{figure}

An empirical correlation between the 1.4 GHz power of the radio relic and the cluster mass has been investigated by \citet{2014MNRAS.444.3130D}, and further revised by \citet{2017MNRAS.472..940K}. 
We show the A2384 radio relic along with the other single and double relics in Fig.~\ref{p1.4vsmass}. This plot can be used to explore the connection between the cluster merger energy dissipation and the non-thermal emission produced by the merger shocks. The double radio relics are found to have their radio powers scaled with the host cluster mass in the form $P_{1.4\mathrm{GHz}} \propto \mathrm{M}^{2.8\pm0.4}$. Single relics have a wider distribution of radio power for any given host cluster mass when compared with that of the double relics.  
The cluster with the lowest host cluster mass in this plane is Abell 168 \citep{2018MNRAS.477..957D}. A2384 with a mass of $M_{500}$ =  2.61 $\times$ 10$^{14}$ M$_{\odot}$, is found to be the second-lowest among the single relic clusters, having a mass very similar to that of the cluster PLCKG200.9-28.2 \citep{2017MNRAS.472..940K}. There are only two double relic clusters with host masses lower than the of A2384, namely, Abell 3365 \citep{2011A&A...533A..35V} and Abell 3376 \citep{2012MNRAS.426.1204K}. The radio power of the A2384 relic is found to be on the scaling relation based on double radio relics.
The single relic symbol sizes are scaled according to their largest linear sizes, and the colour indicates the redshift (Fig.~\ref{p1.4vsmass}). The more powerful single radio relics are also found to be larger in size, but no systematic scaling relation is found. 

\section{Conclusion}
\par In this paper, we show the discovery of a single radio relic associated with the low-mass, bimodal and merging galaxy cluster A2384. In earlier low-frequency GMRT data, this relic was not detected to its full extent, and was confused with the southern radio lobe of the FR I cluster member radio galaxy. In newer sensitive MeerKAT data, we have detected a few new diffuse radio sources, but we need further low-frequency radio and deep X-ray data to characterise these sources and the merger-shock properties. Below, we summarise our main findings: \\
(1) The largest linear size of the newly detected relic is $\sim$ 800 kpc. It is situated perpendicular to the merger axis, and could be a result of the interaction between the subclusters of A2384 after the first passage in the post-merger scenario. The relic is a very steep spectrum source, between 941-1454 MHz, its spectral index $\alpha$ = -2.5 $\pm$ 0.23, suggesting the re-acceleration of the pre-relativistic electrons in the presence of the merger shock. \\ 
(2) In the MeerKAT data, no southern radio lobe of the FR I radio galaxy has been detected, and hence it is not possible to establish any link between the radio galaxy and relic. The available MeerKAT data are not enough to study the link between the FR I galaxy and relic in the presence of the bright radio source  located at the bottom of the relic.\\
(3) We found a candidate radio ridge in the north of the northern radio lobe of the FR I galaxy. The  source is small and is only detected to its full extent in the lowest-frequency band of the MeerKAT. The spectral index of the candidate radio ridge between 941-1283 MHz is $\alpha$ $\sim$ -1.88. Detection of the radio ridge in the MeerKAT observation could be a new class of radio source situated between the two clusters in the hot and dense X-ray filament of A2384. \\

The case of A2384 provides a good opportunity to investigate the possible role of the FR I (LEDA 851827) radio galaxy in generating both the radio relic (via the southern lobe) and the candidate radio ridge (via the northern lobe). Future MeerKAT observations with longer integration time will unveil the nature of the relic and candidate radio ridge. \\
\\
\\
{\it Acknowledgements.} The financial assistance of the South African Radio Astronomy Observatory (SARAO) towards this research is hereby acknowledged. The MeerKAT telescope is operated by the South African Radio Astronomy Observatory, which is a facility of the National Research Foundation, an agency of the Department of Science and Innovation. This work is based upon research supported by the South African Research Chairs Initiative of the Department of Science and Technology and National Research Foundation. (Part of) the data published here have been reduced using the CARACal pipeline, partially supported by ERC Starting grant number 679629 ``FORNAX", MAECI Grant Number ZA18GR02, DST-NRF Grant Number 113121 as part of the ISARP Joint Research Scheme, and BMBF project 05A17PC2 for MeerKAT. Information about CARACal can be obtained online under the URL: https://caracal.readthedocs.io. KT acknowledges support from the Inter-University Institute for Data Intensive Astronomy (IDIA). RK acknowledges the support of the Department of Atomic Energy, Government of India, under project no. 12-R\&D-TFR-5.02-0700 and support from the DST-INSPIRE Faculty Award of the Government of India. DK acknowledges funding from the European Research Council (ERC) under the European Union's Horizon 2020 research and innovation programme (grant agreement no. 679627).\\
{\it Data Availability.} The data underlying this article are subject to an embargo. Once the embargo expires the data will be available (https://archive.sarao.ac.za/; https://www.sarao.ac.za/wp-content/uploads/2019/12/MeerKAT-Telescope-and-Data-Access-Guidelines.pdf).

\bibliography{references}


\end{document}